\documentclass[12pt]{article}

\usepackage[a4paper,left=30mm,right=20mm,top=20mm,bottom=20mm]{geometry}
\usepackage{graphics}
\usepackage{amsmath}
\usepackage{epsfig}
\usepackage{cite}
\usepackage{xcolor}

\newcommand{\bea}{\begin{eqnarray}}
\newcommand{\eea}{\end{eqnarray}}
\newcommand{\be}{\begin{equation}}

\newcommand{\ee}{\end{equation}}

\title{Linearly polarized gluon density at low $x$}

\author{N.A.~Abdulov$^{1}$, X. Chen$^{2,3}$, A.V.~Kotikov$^{1}$, A.V.~Lipatov$^{1,4}$}

\begin{document}

\maketitle

\begin{center}
  {\it $^{1}$Joint Institute for Nuclear Research, 141980, Dubna, Moscow region, Russia}\\
  {\it $^{2}$Institute of Modern Physics, Chinese Academy of Sciences, Lanzhou 730000, China}\\
{\it $^{3}$School of Nuclear Science and Technology, University of Chinese Academy of Sciences, Beijing 100049, China}\\
  {\it $^{4}$Skobeltsyn Institute of Nuclear Physics, Lomonosov Moscow State University, 119991, Moscow, Russia}

\end{center}

\vspace{0.5cm}

\begin{center}

{\bf Abstract }

\end{center}

\indent
Based on a solution of DGLAP evolution equations at small values of Bjorken variable $x$, a simple analytic
approximation for so-called linearly polarized gluon density
${h}_1^{\perp g}(x,k_t^2, Q^2)$  is obtained at LO of perturbation theory.

\vspace{1.0cm}


\newpage

\section{Introduction} \indent

Determination of parton (quark and gluon) distribution functions (PDFs) in a proton 
is a rather important task for modern high energy physics. 
In particular, detailed knowledge on the gluon densities 
is necessary for experiments planned at the Large Hadron Collider (LHC) and future colliders, such as
Electron-Ion Collider (EIC), Future Circular hadron-electron Collider
(FCC-he), Electron-Ion Collider in China (EicC)
and Nuclotron-based Ion Collider fAcility (NICA)\cite{Anderle:2021wcy,Chen:2020ijn,Arbuzov:2020cqg,Abir:2023fpo,SPD}.
For an unpolarized proton, there are distribution
of unpolarized gluons denoted as $f_g(x,Q^2)$ and distribution of linearly polarized gluons
$h_g(x,k_t^2, Q^2)$\cite{Mulders:2000sh,Boer:2011kf},
that corresponds to interference between $\pm 1$ gluon helicity states\footnote{In the literature, other notations $f_1^g(x, Q^2)$ and $h_1^g(x,k_t^2,Q^2)$ or $h_1^{\perp g}(x,k_t^2,Q^2)$ 
are also widely used.}.
The latter is poorly known at the moment\footnote{A theoretical upper bound has been derived, see\cite{Mulders:2000sh,Boer:2011kf}.}
and depends on the gluon transverse momentum $k_t$ (so called Transverse Momentum Dependent, or TMD gluon density).
Of course, it only appears when the gluon $k_t$ becomes non-zero.

One of most rigorous approaches where the TMD quark and gluon densities are involved
is the TMD factorization for transverse momentum  $q_T$ spectra\cite{Collins:1984kg}.
This is an important part of Collins-Soper-Sterman (CSS) resummation\cite{CSS1, CSS2},
which allows one to resum logarithmically
enhanced contributions proportional to $\alpha_s^n \ln^n M/q_T$ to all orders in the QCD coupling.
In this way, one can reproduce the low-$q_T$ behavior of measured transverse momentum spectra of a particle having mass $M$.
Within the TMD factorization, 
effects of linear gluon polarization in transverse momentum spectra of 
Higgs bosons, heavy quarkonia, heavy quark pairs and dijets produced in unpolarized $pp$ collisions have been
studied \cite{Sun:2011iw,Scarpa:2019fol,Boussarie:2023izj}.
It was argued that future LHC data on $\chi_b$ and $\eta_b$ production as well as 
DIS data on heavy quark pair or dijet production expected at FCC-he and EiCC are promising and may exhibit
large $h_g(x,k_t^2,Q^2)$ effects, allowing ways to study its small-$x$ behavior and even saturation dynamics.
However, theoretical predictions are very sensitive to the 
treatment of small and large $b_t$ regions \cite{Boussarie:2023izj}, where $b_t$ is the impact parameter.
In the present note we concentrate mainly of the small-$b_t$ region
and, for the first time, analytically derive the corresponding small-$x$ asymptotic of $h_g(x,k_t^2,Q^2)$.

\section{Approach} \indent

At the leading order (LO) of perturbation theory, gluon densities  at low $k_t$ have the following form\cite{Sun:2011iw}:
\bea
&&{f}_g(x,Q^2) \equiv x {f}^g_1(x,Q^2)
\,, \label{f}\\
&&{h}_g(x,Q^2) \equiv \frac{xk_t^2}{2M^2}\, {h}^{\perp g}_1(x,k^2_t,Q^2)
=\frac{2a_s(Q^2)}{\pi M^2} \int_x^1 \frac{dx_1}{x_1}\left(1-\frac{x}{x_1}\right)
\Bigl[C_Af_g(x_1,Q^2) + \nonumber \\
  &&+C_F f_q(x_1,Q^2)\Bigr]  + ...
\,, \label{h}
\eea
where $C_A=N_c$, $C_F=(N_c^2-1)/(2N_c)$ for the color $SU(N_c)$ group. 
Here
$M=0.938$ GeV is the proton mass,
\be
a_s(Q^2)=\frac{\alpha_s(Q^2)}{4\pi}= \frac{1}{\beta_0\ln(Q^2/\Lambda^2_{\rm QCD})}
\label{as}
\ee
is related to the conventional strong coupling constant $\alpha_s(Q^2)$ and $\beta_0=11-2f/3$ is the first coefficient of QCD
$\beta$-function and $f$ is the number of active quarks.

At the LO and in the limit of small $k_t$ values,
function ${h}_g(x,Q^2)$ is fully determined by the
conventional gluon and quark distributions, $f_g(x,Q^2)$ and $f_q(x,Q^2)$, see~(\ref{h}). The contribution of ${f}_g(x,Q^2)$
is mainly important for small values of the Bjorken variable $x$ but it is restricted as\cite{Mulders:2000sh,Boer:2011kf}
\be
|{h}_g(x,Q^2)| \leq {f}_g(x,Q^2)\,.
\label{Restri}
\ee
So, we will concentrate to this range.

\subsection{Small $x$ asymptotics for  $f_g(x, Q^2)$ and $f_q(x,Q^2)$} \indent

The small-$x$ asymptotic expressions for sea quark and gluon densities $f_a(x, Q^2)$ (where $a = q$ or $g$)
in the framework of so-called generalized {\it double asymptotic scale} (DAS) approach\cite{BF1,Q2evo}
(see also\cite{Rujula})
can be written as follows (both the LO and NLO results and their derivation can be found
\cite{Q2evo,Cvetic1}):
\begin{eqnarray}
f_a(x,Q^2) &=&
f_a^{+}(x,Q^2) + f_a^{-}(x,Q^2), \nonumber \\
f^{+}_g(x,Q^2) &=&
A_g^+	\overline{I}_0(\sigma) \; e^{-\overline d_{+} s} + O(\rho),~~ A_g^+= A_g + \frac{4}{9}
\,A_q,~~
\nonumber \\
f^{+}_q(x,Q^2) &=& A_q^+
\tilde{I}_1(\sigma)  \; e^{-\overline d_{+} s}
+ O(\rho),~~A_q^+= \frac{f}{9}
\, A_g^+,~~
\nonumber \\
f^{-}_g(x,Q^2) &=& A_g^{-} \,
e^{- d_{-} s} \, + \, O(x),~~ A_g^- =- \frac{4}{9}
\,A_q,
        \nonumber \\
	f^{-}_q(x,Q^2) &=&  A_q e^{-d_{-} s} \, + \, O(x),~~ A_a=f_a(x,Q_0^2),
	\label{8.02}
\end{eqnarray}
\noindent
where
$\overline{I}_{\nu}(\sigma)$ and
$\tilde{I}_{\nu}(\sigma)$ ($\nu=0,1$)
are the combinations of the modified Bessel functions $I_{\nu}(\sigma)$
(at $s\geq 0$, i.e. $\mu^2 \geq Q^2_0$) and usual
Bessel functions $J_{\nu}(\tilde{\sigma})$ (at $s< 0$, i.e. $Q^2 < Q^2_0$):
\be
\tilde{I}_{\nu}(\sigma) =
\left\{
\begin{array}{ll}
\rho^{\nu} I_{\nu}(\sigma) , & \mbox{ if } s \geq 0; \\
(-\tilde{\rho})^{\nu} J_{\nu}(\tilde{\sigma}) , & \mbox{ if } s < 0,
\end{array}
\right. \, ~~
\overline{I}_{\nu}(\sigma) =
\left\{
\begin{array}{ll}
\rho^{-\nu} I_{\nu}(\sigma) , & \mbox{ if } s \geq 0; \\
\tilde{\rho}^{-\nu} J_{\nu}(\tilde{\sigma}) , & \mbox{ if } s < 0.
\end{array}
\right. 
\label{4}
\ee
Functions $I_{\nu}(\sigma)$ and  $J_{\nu}(\sigma)$ have the following series representations:
\be
I_{\nu}(\sigma)=\sum_{m=0}^{\infty} \, \frac{1}{k!(k+\nu)!}\, \left(\frac{\sigma}{2}\right)^{2k+\nu},~~
J_{\nu}(\sigma)=\sum_{m=0}^{\infty} \, \frac{(-1)^k}{k!(k+\nu)!}\,  \left(\frac{\sigma}{2}\right)^{2k+\nu}\,.
\label{4a}
\ee
\noindent
We note that $\overline{I}_{0}(\sigma) = \tilde{I}_{0}(\sigma)$ and
\bea
&&s=\ln \left( \frac{a_s(Q^2_0)}{a_s(Q^2)} \right),~~
\sigma = 2\sqrt{\left|\hat{d}_+\right| s
  \ln \left( \frac{1}{x} \right)},~~ \nonumber \\
&&\rho=\frac{\sigma}{2\ln(1/x)},~~
\tilde{\sigma} = 2\sqrt{-\left|\hat{d}_+\right| s
  \ln \left( \frac{1}{x} \right)},~~ \tilde{\rho}=\frac{\tilde{\sigma}}{2\ln(1/x)},
\label{intro:1a}
\eea
and
\begin{equation}
  \hat{d}_+ =
  - \frac{12}{\beta_0},~~~
  \overline d_{+} =
1 + \frac{20f}{27\beta_0},~~~
d_{-} =
\frac{16f}{27\beta_0}
\label{intro:1b}
\end{equation}
are the singular and regular parts of the anomalous dimensions.

\subsection{Small $x$ asymptotics for ${h}_g(x,Q^2)$ } \indent

Now, using the results (\ref{8.02}) we can evaluate the r.h.s. in (\ref{h}) and derive analytical expression for linearly polarized
gluon density ${h}_g(x,Q^2)$, which
has the following form:
\begin{eqnarray}
{h}_g(x,Q^2) &=&
h_g^{+}(x,Q^2) + h_g^{-}(x,Q^2), \nonumber \\
h^{+}_g(x,Q^2) &=& \frac{2a_s(Q^2)}{\pi M^2}
\left\{C_AA_g^+ \left(\overline{I}_1(\sigma)-\overline{I}_0(\sigma)\right)
+C_FA_q^+ \left(\overline{I}_0(\sigma)-\tilde{I}_1(\sigma)\right) \right\}\; e^{-\overline d_{+} s},~~
 \nonumber \\
h^{-}_g(x,Q^2) &=& 0\,.
	\label{8.02h}
\end{eqnarray}
\noindent
Thus, at low $x$ range the ${h}_g(x,Q^2)$ rises faster than $f_g(x,Q^2)$ since
$\overline{I}_1(\sigma)/\overline{I}_0(\sigma) \sim \rho^{-1} \sim \sqrt{x}$, that is illustrated below.
Note that in the wider $x$ range one can use the following extension
(see \cite{Kotikov:2017mhk,Kotikov:2021yzb}):
\be
f_g^{\pm}(x,Q^2) \to f_g^{\pm}(x,Q^2) (1-x)^{b_{\pm}},~~ h_g^{\pm}(x,Q^2) \to h_g^{\pm}(x,Q^2) (1-x)^{b_{\pm}},
	\label{8.02h1}
\ee
where $b_{-} \sim 4$ and $b_{+} \sim 5$, coming from quark counting rules \cite{Matveev:1973ra}.

\section{Results} \indent

\begin{figure}
\centering
\vskip 0.5cm
\includegraphics[width=7.9cm]{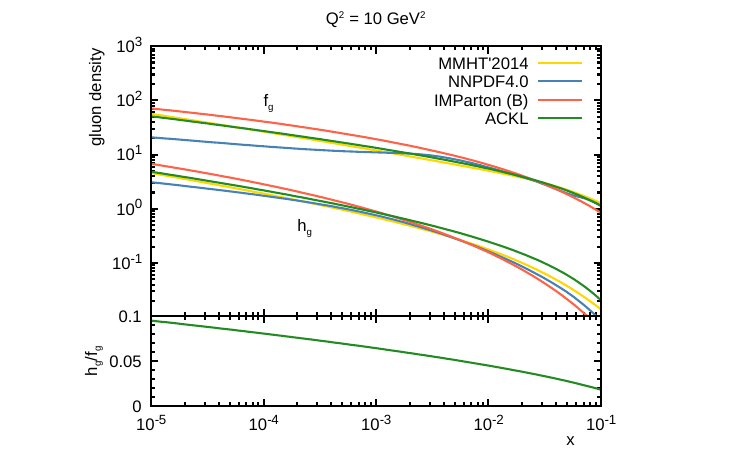}
\includegraphics[width=7.9cm]{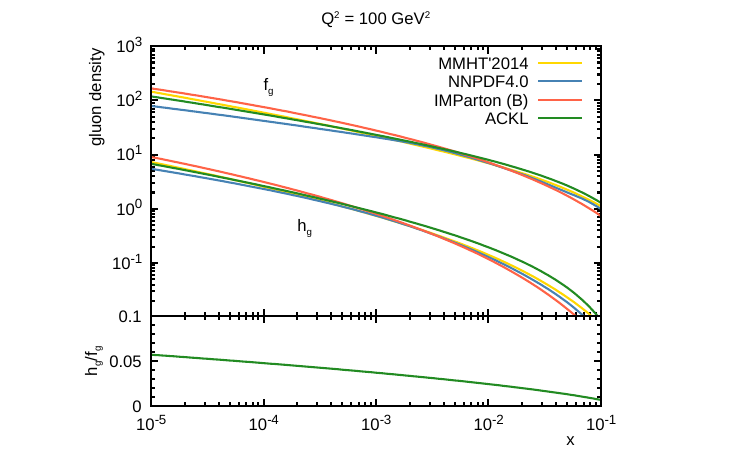}
\includegraphics[width=7.9cm]{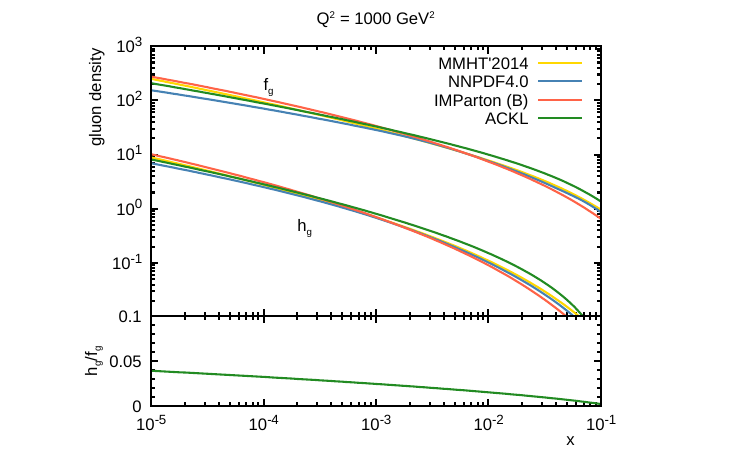}
\includegraphics[width=7.9cm]{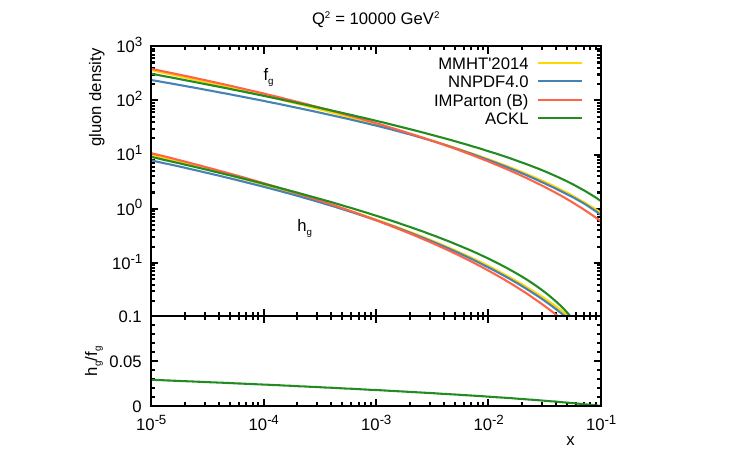}
\vskip -0.3cm
\caption{$x$ dependence of $h_{g}(x,Q^2)$ (below curves), $f_{g}(x,Q^2)$ (upper curves) and their ratio
  for varions $Q^2$ values.
}
\end{figure}

Now we can present numerical results for linearly polarized gluon density ${h}_g(x,Q^2)$ in a proton at low $x$.
Here we
employ the values of phenomenological parameters $A_g$, $A_q$ and $Q_0^2$ which were extracted
from the experimental data on proton structure function $F_2(x,Q^2)$\cite{Cvetic1}.
Our predictions are shown in Fig. 1,
where we plot both ${h}_g(x,Q^2)$ and $f_g(x, Q^2)$
as well as their ratio $r_h(x,Q^2)={h}_g(x,Q^2)/f_g(x,Q^2)$
for different $Q^2$ values.
Here we use derived analytical expressions~(\ref{8.02}) and (\ref{8.02h})
and perform their comparison with numerical calculation of the
integral representation (\ref{h}),
where three different PDF sets, namely, MMHT'2014 (LO)\cite{MMHT}, IMParton (B) \cite{Wang:2016sfq}
and more recent NNPDF4 (LO)\cite{NNPDF} have been applied.
One can see that our analytical results are in good agreement with corresponding numerical calculations,
especially with the MMHT'2014 (LO) set.
We find that linearly polarized gluon density
${h}_g(x, Q^2)$ rises faster at $x \to 0$ then the conventional gluon distribution $f_{g}(x,Q^2)$. This rise is clearly seen
from the ratios $r_h(x, Q^2)$, shown also in Fig.~1.

Note also that in a recent paper\cite{EMCh} we studied modifications of the linearly polarized gluon density
${h}_1^{\perp g}(x,k_t^2,Q^2)$ in the nuclei targets. Within the simplest 
rescaling model\cite{Close:1983tn,Close:1984zn}, extended to the low $x$ range\cite{Kotikov:2017mhk},
shadowing and antishadowing effects for ${h}_1^{\perp g}(x,k_t^2,Q^2)$ look very similar to those in the case of
$f_g(x, Q^2)$, observed in\cite{Abdulov:2022ypq}.

\section{Conclusion} \indent

At the leading order of perturbation theory,
we derived a simple analytical expression for small-$x$ asymptotics of 
linearly polarized gluon density in a proton ${h}_1^{\perp g}(x,k_t^2, Q^2)$ 
and studied its behavior 
within the framework of the generalized DAS approach.
We found that linearly polarized gluon distribution
rises faster 
than the usual one $f_g(x,Q^2)$.
Our analytical results are in good agreement with the numerical calculations based on 
the MMHT'2014 (LO), IMParton (B) and NNPDF4.0 (LO) sets.
The derived expressions could be useful 
for forthcoming phenomenological studies. \\

Researches described in Section~2 were 
supported by the Russian Science Foundation under grant 22-22-00387.
Studies described and performed in Sections 3 and 4 were supported by the Russian Science Foundation under grant 22-22-00119.

\end{document}